\documentclass[letterpaper]{article} 
\usepackage{aaai24}
\usepackage{times}  
\usepackage{helvet}  
\usepackage{courier}  
\usepackage[hyphens]{url}  
\usepackage{graphicx} 
\usepackage{natbib}  
\usepackage{caption} 
\usepackage{algorithm}
\usepackage{graphicx}
\usepackage{amsmath}
\usepackage{algpseudocode}
\usepackage{amsmath,amssymb,amsfonts}
\usepackage[capitalise]{cleveref}
\usepackage{subcaption}
\usepackage{makecell}
\usepackage{float}

\usepackage{newfloat}
\usepackage{listings}
\DeclareCaptionStyle{ruled}{labelfont=normalfont,labelsep=colon,strut=off} 
\floatstyle{ruled}
\newfloat{listing}{tb}{lst}{}
\floatname{listing}{Listing}
\title{CoSS: Co-optimizing Sensor and Sampling Rate for Data-Efficient AI in Human Activity Recognition}
\author{
    Mengxi Liu\textsuperscript{\rm 1}, Zimin Zhao \textsuperscript{\rm 2}, Daniel Geißler\textsuperscript{\rm 1}, Bo Zhou\textsuperscript{\rm 1,2}, Sungho Suh\textsuperscript{\rm 1,2}, Paul Lukowicz\textsuperscript{\rm 1,2}
}
\affiliations{
    \textsuperscript{\rm 1}German Research Center for Artificial Intelligence (DFKI), Kaiserslautern, Germany\\
    \textsuperscript{\rm 2}Department of Computer Science, RPTU Kaiserslautern-Landau, Kaiserslautern, Germany\\
    
    Email: firstname.lastname@dfki.de
}

\usepackage{bibentry}
\begin{document}

\maketitle

\begin{abstract}

Recent advancements in Artificial Neural Networks have significantly improved human activity recognition using multiple time-series sensors.
While employing numerous sensors with high-frequency sampling rates usually improves the results, it often leads to data inefficiency and unnecessary expansion of the ANN, posing a challenge for their practical deployment on edge devices. 
Addressing these issues, our work introduces a pragmatic framework for data-efficient utilization in HAR tasks, considering the optimization of both sensor modalities and sampling rate simultaneously. 
Central to our approach are the designed trainable parameters, termed 'Weight Scores,' which assess the significance of each sensor modality and sampling rate during the training phase. 
These scores guide the sensor modalities and sampling rate selection. 
The pruning method allows users to make a trade-off between computational budgets and performance by selecting the sensor modalities and sampling rates according to the weight score ranking.
We tested our framework's effectiveness in optimizing sensor modality and sampling rate selection using three public HAR benchmark datasets. 
The results show that the sensor and sampling rate combination selected via CoSS achieves similar classification performance to configurations using the highest sampling rate with all sensors, but at a reduced hardware cost.


\end{abstract}

\section{Introduction}
Human Activity Recognition (HAR) based on Artificial Neural Networks (ANN) has achieved great success in the past years in many application areas like healthcare \cite{karar2022survey}, fitness \cite{phukan2022convolutional}, smart home \cite{bianchi2019iot} and smart manufacturing \cite{suh2023worker}. 
The concurrent use of multiple sensors for human activity recognition provides a practical solution for complex activity recognition while improving recognition accuracy, robustness, and reducing noise \cite{aguileta2019multi}. 
In addition, human activities consist of complex sequences and motor movements, whereas capturing these temporal dynamics is fundamental for successful human activity recognition (HAR) \cite{ordonez2016deep}. 
The high sampling rates can often contribute to better results, especially for complex activity recognition. 
However, the multiple-modality sensor-based solution and unnecessarily high sampling rate often require a larger Artificial Neural Networks (ANN) model for information fusion and higher power consumption for data acquisition and transmission, resulting in a challenging task towards practical deployment on edge devices where power conservation is crucial.
While advancements in neural network compression, ranging from pruning \cite{han2015learning} and quantization \cite{nagel2021white} to knowledge distillation \cite{hinton2015distilling}, have facilitated lightweight inference models on edge devices, the efficient utilization of sensor data for comprehensive AI model optimization, including sensor modalities and sample rates, remains a gap in current literature.

The purpose of sensor optimization is to remove sensors with minimal impact on classification results. 
Thus, the sensors’ optimization can produce the beneﬁt to reduce the input and computational complexity, while preserving the essential accuracy for HAR \cite{espinilla2017optimizing}.
Although reduced sampling rates imply more efficient resource utilization in real-world wearable HAR systems, existing works have predominantly overlooked the co-optimization of sensor modality selection and sampling rates.

When multiple sensors are used simultaneously, the sampling rate may vary between sensors depending on their nature and the type of data they capture. 
Optimizing these sampling rates involves not only considering each sensor individually but also how their data complements each other. 
Thus, we propose CoSS, a general framework focused on co-optimizing sensor modality and sampling rate for data-efficient AI in HAR tasks simultaneously requiring only training once.
With CoSS, the trainable parameters "weight score" were designed to evaluate the importance of each sensor and sampling rate during training. 
The weight score ranking directs the selection of sensor modalities and sampling rates after training according to the trade of performance and hardware cost.

The main contributions of this paper can be summarized as follows.
\begin{itemize}
    \item We propose a general efficient framework CoSS to address co-optimizing sensor and sampling rate for data-efficient AI in HAR tasks simultaneously requiring only training once.
    \item We demonstrate our framework's effectiveness in optimizing sensor modality and sampling rate selection using three public HAR benchmark datasets. 
\end{itemize}

\section{Related work}

\subsection{Sampling rate optimization}

Research in HAR has explored various methods to optimize sampling rates, including adaptive frameworks \cite{yang2023freqsense}, deep learning models \cite{qi2013adasense}, and sensor-specific algorithms \cite{anish2019sensor}. These approaches aim to achieve a balance between accuracy and efficiency in HAR systems. 
For instance, FreqSense \cite{yang2023freqsense}, an adaptive resolution network, employed a subsampling strategy with conditional early exits, allowing variable resolution processing for different sample complexities. 
Although effective in balancing computational efficiency and accuracy, these networks are larger than typical neural networks, posing challenges for deployment on memory-limited edge devices. 
Additionally, while they reduce network complexity for simpler activities, their power consumption during data acquisition remains unchanged due to the need for high sampling rates for complex activities. 
Another proposed solution involves sensor-classifier co-optimization \cite{anish2019sensor}, but this adds hardware complexity and potential power usage. 
Hence, there is a gap in research regarding optimal overall sampling rates without requiring additional sensors. 



\subsection{Sensor optimization}

The sensor optimization solution in the existing works includes the exhaustive search \cite{ertuǧrul2017determining,aziz2016identifying}, feature or classification selection \cite{leite2021optimal,tian2020optimizing} and adaptive-context-aware \cite{zappi2008activity}, these methods often suffer from high computational cost and time-consuming at training phase.
For instance, an exhaustive search method was applied in the work \cite{aziz2016identifying} to identify the number and location of body-worn sensors for classifying walking, transferring, and sedentary activities; the computational cost of such a solution will be increased exponentially along with the number of the sensors to be optimized.
Besides, this proposed method can not provide a ranking of the importance of the sensor modalities to guide users in selecting the sensor flexibly according to the hardware budgets. 
The proposed CoSS framework, while similar to Leite's work in sensor selection, introduces a more efficient approach. Leite's method uses a lightweight neural network and a sensor channel selection algorithm to rank and eliminate less important channels, requiring multiple training sessions, which is time and energy-intensive. In contrast, CoSS needs only a single training session, leveraging pruning technology to directly obtain sensor-pruned results. Additionally, while Leite's method evaluates sensor channel importance using $N\times(N-1)/2$ trainable weights, CoSS simplifies this with just $N$  trainable weights, enhancing efficiency.
Last but not least, the proposed CoSS framework focuses on the co-optimization of sensors and sampling rate together.


\section{Proposed Framework}
\begin{figure*}[!t]
\footnotesize
\centering
\includegraphics[width=1.0\linewidth]{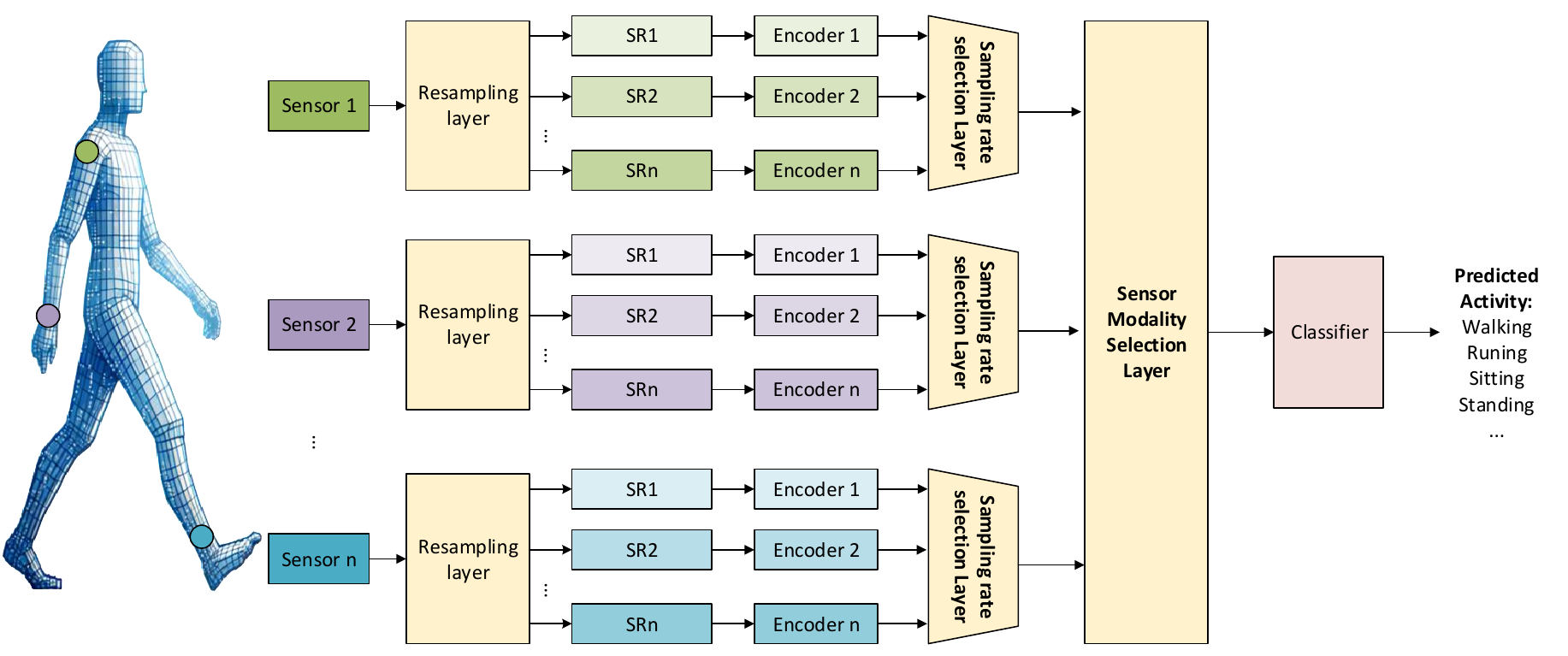}
\caption{Proposed CoSS framework, the framework is based on multi-branch feature fusion architecture, where three additional layers, such as resampling layers, sampling rate, and sensor selection layers, are embedded. \textbf{SRn} is the downsampling data from the resampling layer}
\label{fig:framework}
\end{figure*}

\cref{fig:framework} shows the architecture of the proposed framework CoSS for sensor and sampling rate selection.
The architecture of CoSS is designed based on the feature-level fusion architecture \cite{qiu2022multi}.
Compared to the general feature-level fusion ANN model, three additional layers are integrated into the ANN model in CoSS architecture, such as resampling layers, sampling rate selection layers, and sensor selection layers, by which the importance ranking of sensor and sampling rate can be obtained by comparing the weight scores after training.

\subsection{Resampling layer}
The resampling layers in the system are engineered to produce multiple candidates at varied sampling rates. Each sensor node features a dedicated resampling layer that processes input data from sensors at the original sampling rate. 
This layer cycles through a set list of resampling rates, yielding several branches of down-sampled data. Notably, the sampling rates are highly flexible, allowing for arbitrary configurations, including fractional values.
The resampling step size ($S$), is calculated based on the original sampling rate ($f_{original}$), and target sampling rate ($f_{target}$) as shown in \cref{eq: step_size}.
\begin{equation}
    \begin{aligned}
    \label{eq: step_size}
    S = \frac{f_{original}}{f_{target}}
    \end{aligned}
\end{equation}

However, since the step size ($S$) could be a float value, the generated index of original data for resampling can result in a non-integer value, which is illegal. 
To address this, we use linear interpolation to sample the target data (TD) from the original data (OD) according to \cref{eq: downsampling}.
\begin{equation}
    \begin{aligned}
    \label{eq: downsampling}
    TD[i] = OD[\lfloor i \times S \rfloor] + (i\times S - \lfloor i \times S \rfloor) \times\\
    (OD[\lceil i \times S \rceil] - OD[\lfloor i \times S \rfloor])
    \end{aligned}
\end{equation}
 where $i$ is the new index of the target data range from zero to the length of the new target data.
 
Since the window size of the target data and original data is different after resampling layers, an adaptive kernel size ($KS$) was used in each feature extraction branch to guarantee that filters from different branches process the temporal information with equal time length. The adaptive kernel size is computed as follows.
\begin{equation}
    \begin{aligned}
    \label{eq: kernel_size}
    KS_{target} =\lfloor KS_{original} \times \frac{f_{target}}{f_{original}} \rfloor
    \end{aligned}
\end{equation}
Thus, the sliding windows with a lower sampling rate have a smaller kernel size, leading to a smaller model size and lower computational load.

\subsection{Sampling rate selection layer}

The resampling rate selection layer is designed to rank the importance of different downsampling rates for classification tasks.
We consider there are $M$ sensor modalities and $n_i$ downsampling rate candidates for $sensor_i$.
Each $sensor_i$'s sampling rate selection layer comprises $n_i$ trainable weights $\alpha$, which indicates the importance of specific downsampling sampling rates to the final classification result. 
The extracted feature $Fsr_i$ from the $i_{th}$ sensor with different sampling rates is calculated according to \cref{eq: samplingrate_selection}.
\begin{equation}
    \begin{aligned}
    \label{eq: samplingrate_selection}
    Fsr_i = \sum_{j=1}^{n_i}{F_{i,j} \times \frac{exp(\alpha_j)}{\sum_{k=1}^{n_i}exp{(\alpha_k)}}}
    \end{aligned}
\end{equation}

In the current literature, most of the results demonstrate that the higher sampling rate often leads to a better result.
Thus, importance ranking and sampling rate could be closely related: candidates with higher sampling rates are supposed to have higher weight scores.
During the sampling rate selection, sensors that are less sensitive to changes in the sampling rate are configured at a lower sampling frequency, by which the power consumption could be decreased compared to a uniform sampling rate configuration for all sensors.


\subsection{Sensor selection layer}

Similar to sampling rate selection, we assign $M$ trainable weights $\alpha$ to $M$ sensors, and the mixed features $F_{mix}$ from the sensor selection layer are calculated as \cref{eq: sensorslection} shows.

\begin{equation}
    \begin{aligned}
    \label{eq: sensorslection}
    F_{mix} = \sum_{i=1}^{M}{Fsr_{i} \times \frac{exp(\alpha_i)}{\sum_{k=1}^{M}exp{(\alpha_k)}}}
    \end{aligned}
\end{equation}

Thus, there are $M*N$ trainable weights $\alpha$ for both sampling rate and sensor selections in the proposed CoSS framework. 
The output of the sensor selection layer is fed to the fully connected layers for the final classification.



All trainable weights $\alpha$ can be obtained directly after model training.
A progressive pruning method similar to the work \cite{han2015learning} is applied to remove the sensors with lower weights.
The feature branch with the selected sampling rate data input is kept while the branches, including the rest of the sampling rate data, are pruned.
The model's performance with the chosen sensors and sampling rate can be assessed without retraining the model from the beginning.
In addition, fine-tuning can be used to improve the performance after sensor and sampling rate selection.
In this work, we conducted an additional 10 training epochs for fine-tuning.


\section{Evaluation}
To evaluate our framework’s effectiveness in optimizing sensor modality and sampling rate selection, the experiments on three public HAR benchmark datasets were conducted.

\begin{figure}[t]
\centering
\includegraphics[width=1.0\linewidth]{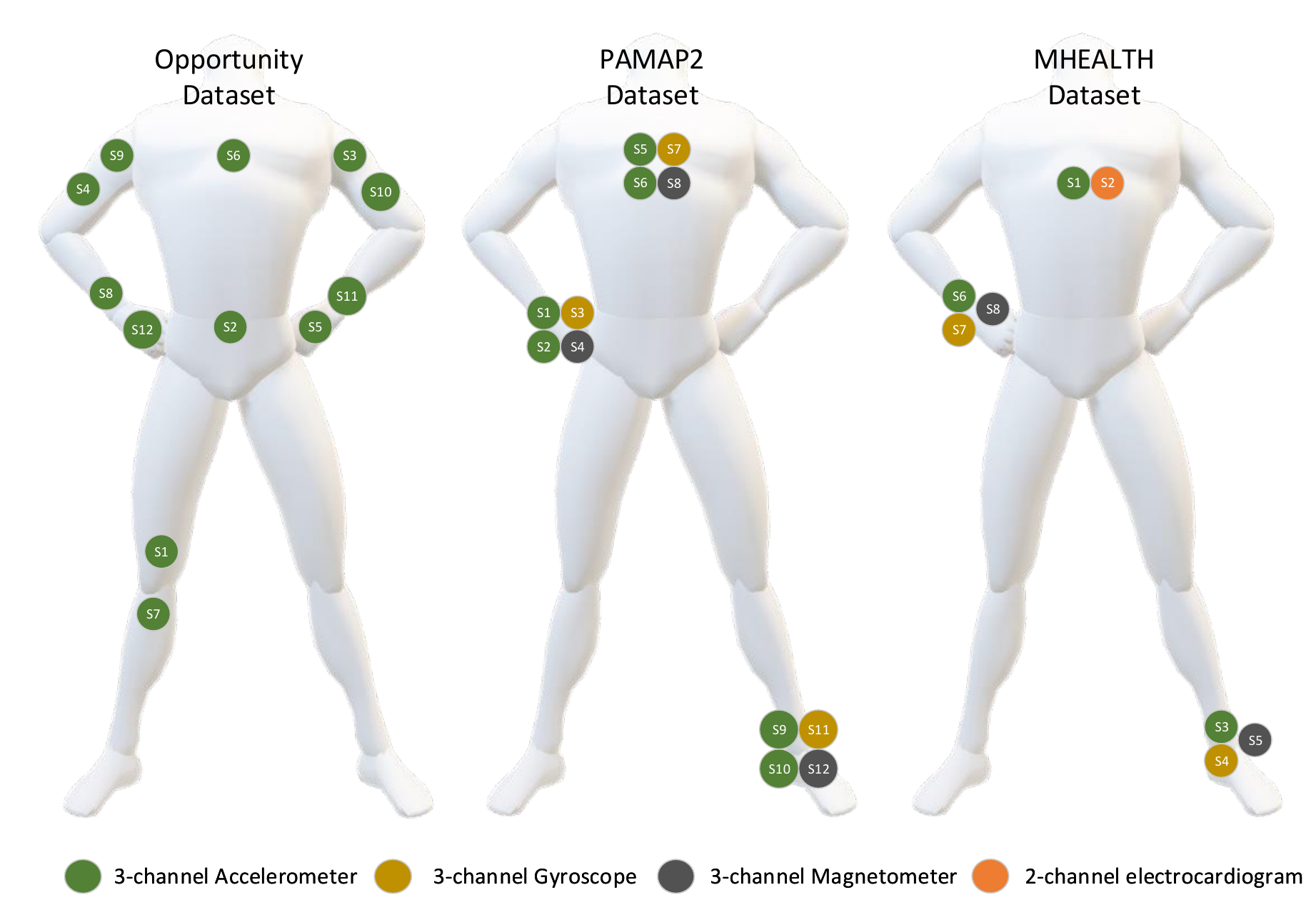}
\caption{Sensor configurations of the three datasets, every sensor is annotated with an identifying number in this work}
\label{fig:DatasetInfo}
\end{figure}

\subsection{Dataset}
\cref{fig:DatasetInfo} shows the sensor configuration of the three datasets.
The Opportunity dataset is a collection of sensor data used for human activity recognition and ambient intelligence research. 
It features a range of sensors with original sampling rate of 30 Hz, like accelerometers and gyroscopes, placed on multiple subjects and in their environment. 
The dataset includes diverse daily activities. There are 5 classes for the locomotion recognition task and 18 classes for the gesture recognition task. Only gesture recognition is tested in this work.
The PAMAP2 dataset is designed for physical activity monitoring research. It comprises sensor data from accelerometers, gyroscopes, and magnetometers placed on the body of participants. The original sampling frequency of this dataset is 100 Hz.
These participants, a diverse group, engaged in a variety of physical activities ranging from simple actions like walking and sitting to more vigorous exercises like running and cycling. 
The MHEALTH dataset is aimed at mobile health monitoring.
It contains sensor data from accelerometers, gyroscopes, and ECG monitors attached to subjects. 
These subjects performed a series of activities, including both basic daily movements and fitness exercises.
The original sampling rate of this dataset is 50 Hz.




Since the purpose of this work is not to discuss the neural network's performance across subjects or sessions, the leave-one-out cross-validation method was not utilized in this work. 
During the experiment, we divided all data from each dataset into three groups randomly: 70 \% training dataset, 15 \% validation dataset, and 15 \% test dataset.

\subsection{ANN Model}
The neural network used for the classification tasks consists of two blocks: encoders and a classifier.
The number of encoders equals $M*N$, $M$ is the number of sampling rates to be evaluated and $N$ is the number of sensors.
The encoders consist of Convolution (CNN), Batch Normalization(BN), and ReLU activation function; the architecture of the encoder is as follows: $CNN-ReLU-BN-CNN-ReLU-BN-CNN-ReLU-Pooling-Flatten$. 
The filter number of each convolutional layer is 100.
The original kernel sizes of the three experiments are 9 (Opportunity), 20 (PAMAP2) and 10 (MHEALTH).
The classifier block consists of two dense layers.
The cross-entropy loss function and SGD optimizer with a learning rate of 1e-2, a weight decay of 0.9, and a momentum of 1e-4, as well as a batch size of 512, were selected to train this model. 
The model was trained for 300 epochs with an early stopping using the patience of 30 to avoid overfitting. 
The ANN model was implemented in PyTorch. 
The experiments conducted on each dataset were repeated ten times to eliminate the possibility of randomness.

\subsection{Experiment Result}

\begin{figure*}[!t]
\footnotesize
\centering
\includegraphics[width=1.0\linewidth]{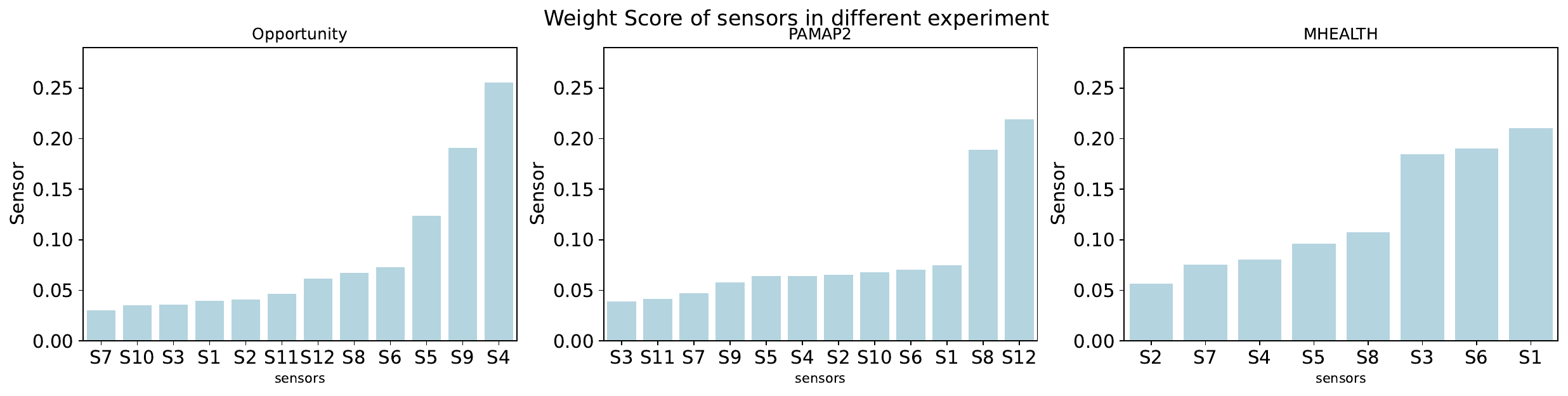}
\caption{Weight score of sensors after training in different experiments}
\label{fig:sensor_weight_results}
\end{figure*}

\begin{figure*}[!t]
\footnotesize
\centering
     \begin{subfigure}[b]{0.33\textwidth}
         \centering
         \includegraphics[width=\textwidth]{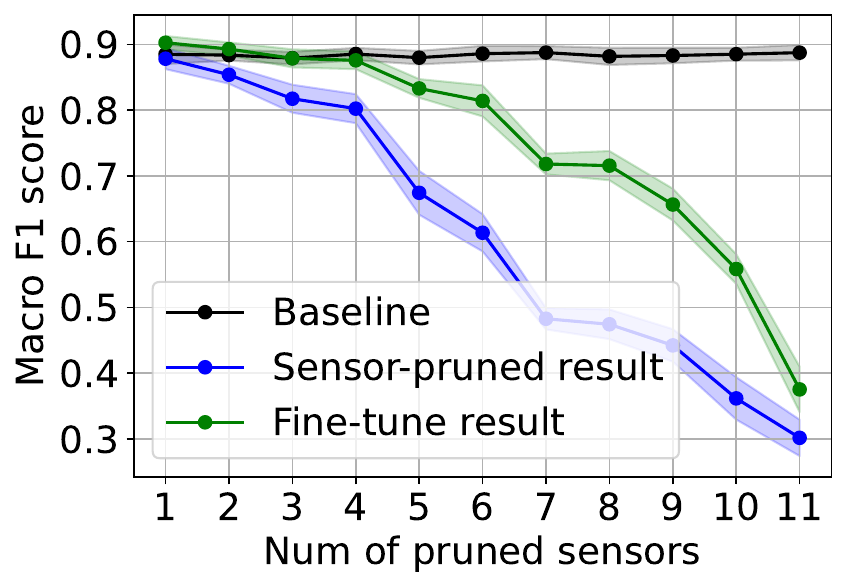}
         \caption{Opportunity-gesture}
         \label{fig:Opportunity_sensor_optimization_result}
     \end{subfigure}
     \hfill
     \begin{subfigure}[b]{0.33\textwidth}
         \centering
         \includegraphics[width=\textwidth]{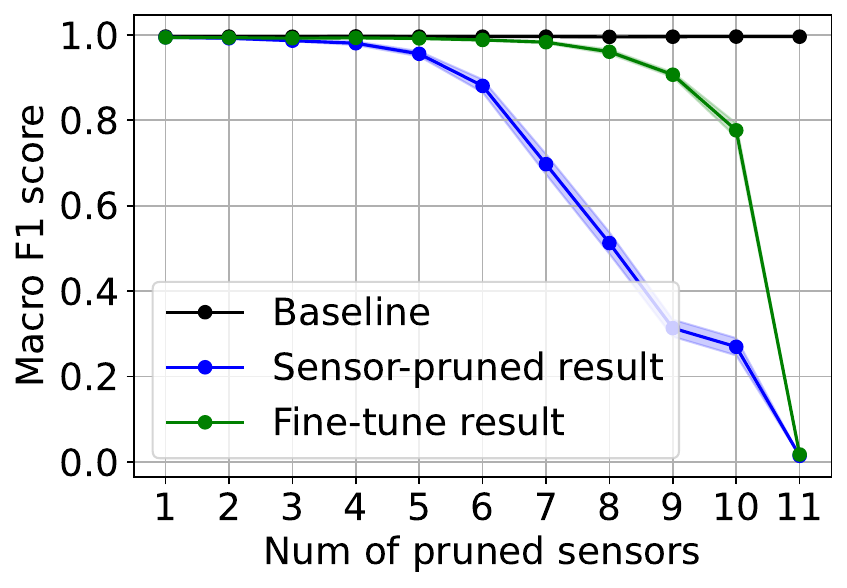}
         \caption{PAMAP2}
         \label{fig:PAMAP2_sensor_optimization_result}
     \end{subfigure}
     \hfill
     \begin{subfigure}[b]{0.33\textwidth}
         \centering
         \includegraphics[width=\textwidth]{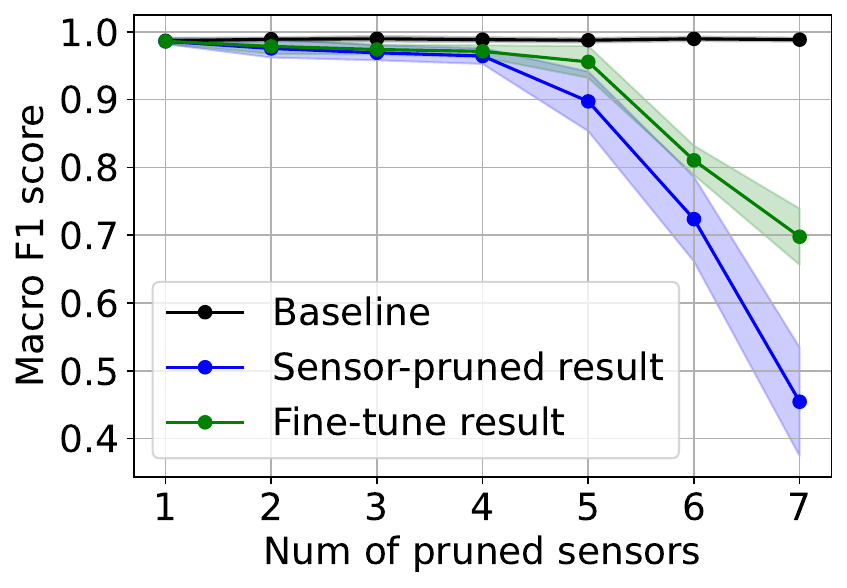}
         \caption{MHEALTH}
         \label{fig:MHEALTH_sensor_optimization_result}
     \end{subfigure}
     
    \caption{Sensor optimization result}
    \label{fig:sensor_optimization_result}
\end{figure*}

\begin{figure*}[!t]
\footnotesize
\centering
\includegraphics[width=1.0\linewidth]{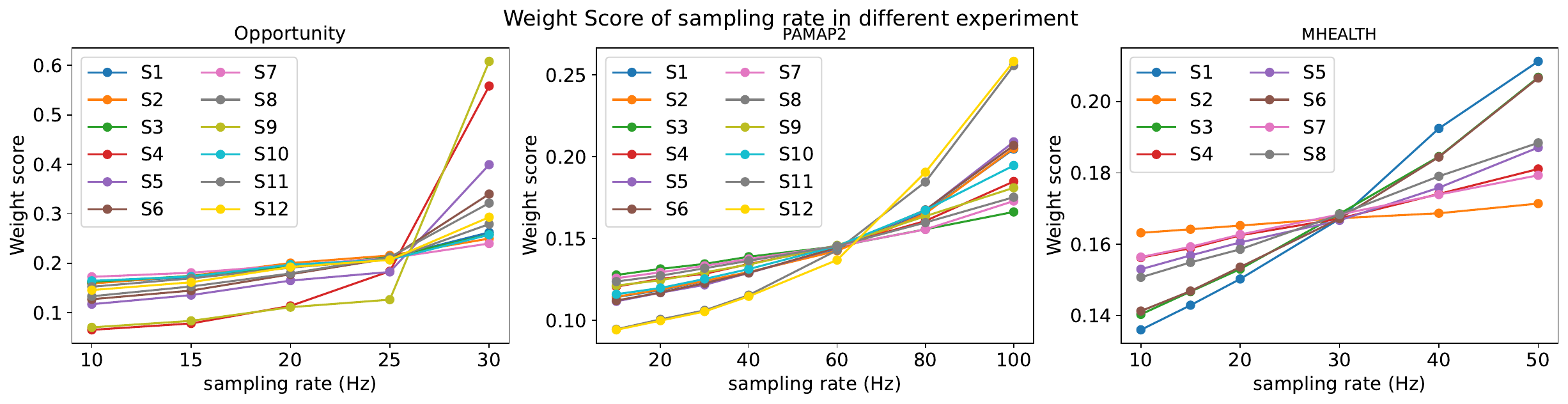}
\caption{Weight score of sampling rate after training in different experiments}
\label{fig:samplingrate_weight_results}
\end{figure*}

\begin{figure*}[!t]
\footnotesize
\centering
     \begin{subfigure}[b]{0.33\textwidth}
         \centering
         \includegraphics[width=\textwidth]{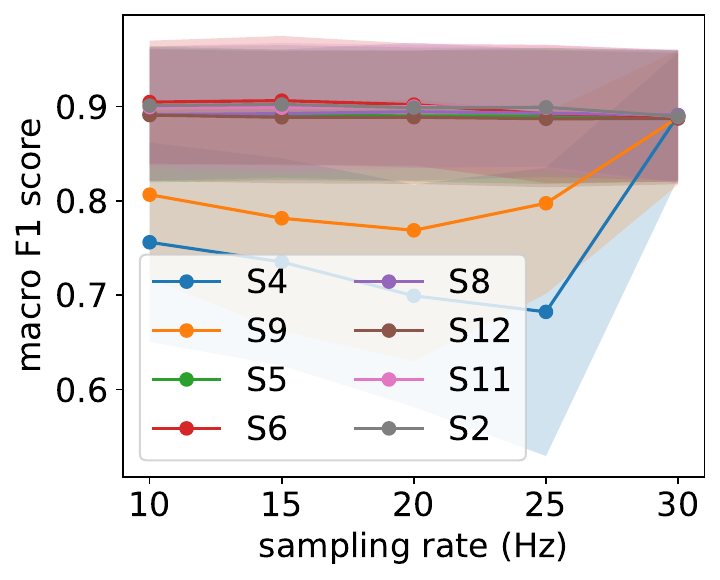}
         \caption{Opportunity-gesture}
         \label{fig:Opportunity_gestures_samplingrate_performance}
     \end{subfigure}
     \hfill
     \begin{subfigure}[b]{0.33\textwidth}
         \centering
         \includegraphics[width=\textwidth]{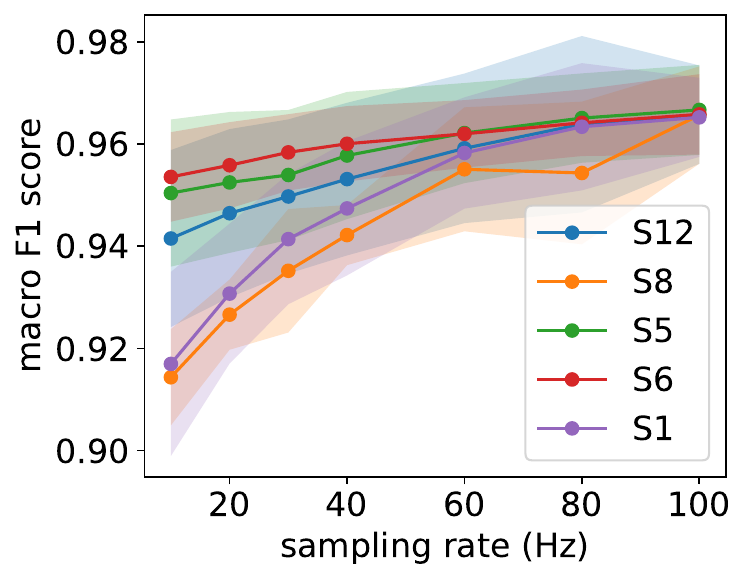}
         \caption{PAMAP2}
         \label{fig:PAMAP2_gestures_samplingrate_performance}
     \end{subfigure}
     \hfill
     \begin{subfigure}[b]{0.33\textwidth}
         \centering
         \includegraphics[width=\textwidth]{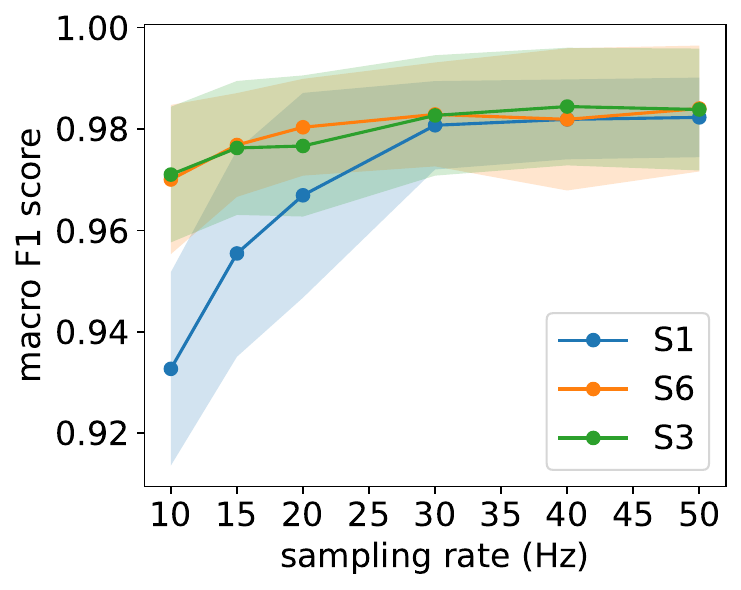}
         \caption{MHEALTH}
         \label{fig:MHEALTH_gestures_samplingrate_performance}
     \end{subfigure}
     
    \caption{The classification performance of the sampling rate selection of the selected sensors (without fine-tuning)}
    \label{fig:Sampling_optimization_result}
\end{figure*}


The importance rankings of sensor modalities and sampling rates are determined after training. Based on these rankings, we can progressively prune less significant sensor modalities and select the most optimal sampling rates.  \cref{fig:sensor_weight_results} displays the sensor weight scores in three distinct experiments, each utilizing different datasets. These results highlight that the weights of various sensors differ significantly; sensors at the lower end of the ranking have scores around 0.05, whereas the top scores in all three experiments approximate 0.20. Subsequently, a progressive pruning approach was employed, removing 
the feature extraction branch of the sensors from the neural networks based on their weight rankings. The performance of the models, post-pruning, was evaluated through direct model inference without the need for training.  \cref{fig:sensor_optimization_result} illustrates these pruning results. We observed that the removal of four sensors from both the PAMAP2 and MHEALTH datasets had a minimal impact on performance. In contrast, the Opportunity dataset showed slight performance alterations due to sensor pruning. However, significant improvements were noted after fine-tuning. The result indicates that four, seven, and five sensors can be removed from the Opportunity-gestures, PAMAP2, and MHEALTH datasets, respectively, without noticeable degradation in classification results post fine-tuning.

\begin{table*}[!t]
\centering
\renewcommand{\arraystretch}{1.2}
\footnotesize
\caption{Result Summary }
\label{tab:result_summary}
    \begin{tabular}{c|c|c|c}
    \hline
    Experiment & Opportunity(gestures)  & PAMAP2 & MHEALTH\\ \hline
    \#Sensor &12& 12& 8\\ \hline
    Sampling rates (Hz)&30,25,20,15,10& 100,80,60,40,30,20,10& 
    50,40,30,20,15,10\\ \hline
    Window size (s)&3& 2& 4\\ \hline
    \multicolumn{4}{l}{Sensor pruned result (with original sampling rate)} \\\hline
    Pruned sensor & S7,S10,S3,S1 &  S3, S11,S7,S9,S5,S4,S2& S2,S7,S4,S5,S8\\ 
    Performance reduction(\%) &0.95 $\pm$ 1.09& 1.30 $\pm$ 0.23& 3.19 $\pm$ 2.23\\ 
     Model size reduction(MB) & 2.81/8.47(33\%) &  4.90/8.46 (58\%)& \textbf{3.51/5.66 (62\%)}\\ \hline
    \multicolumn{4}{l}{Sampling rate selection result (with selected sensors)} \\\hline
    Selected sampling rate (Hz)&\makecell{30(S4),30(S9),10(S5),10(S6),\\10(S8),10(S12),10(S11),10(S2)}&\makecell{80(S12),100(S8), 60(S5),40(S6),\\80(S1)}& 30(S1), 20(S6), 30(S3)\\ 
    Performance reduction(\%)&1.85 $\pm$ 1.01&0.42 $\pm$ 0.22& 0.29 $\pm$ 0.50\\ 
    \hline
    \end{tabular}
\end{table*}

\begin{figure*}[!t]
\footnotesize
\centering
\includegraphics[width=1.0\linewidth]{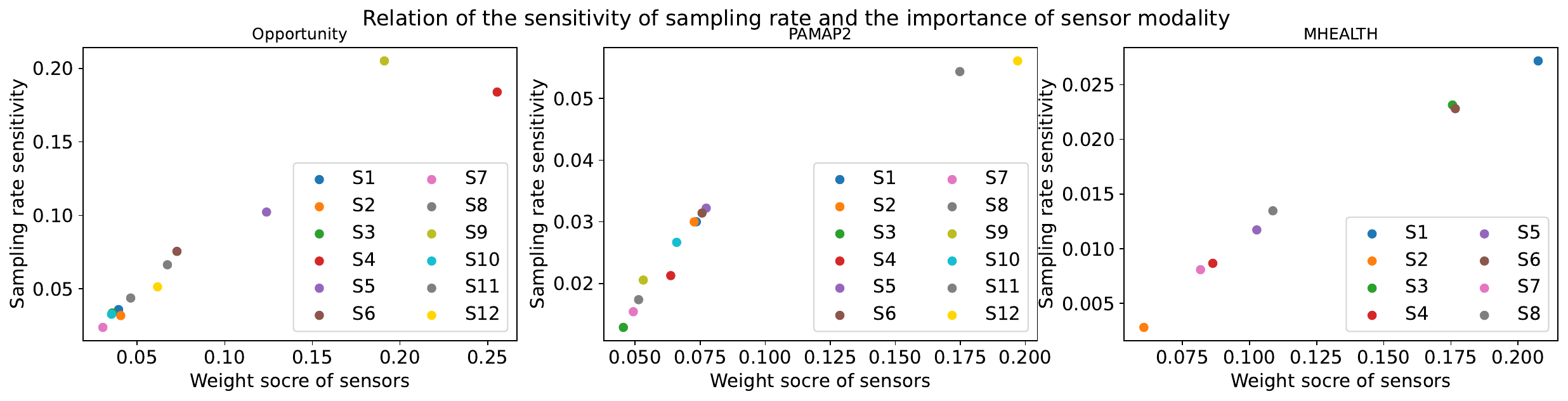}
\caption{The relation of sampling rate sensitivity and the importance of sensor modality}
\label{fig:relation_sensor_samplingrate}
\end{figure*}


\cref{fig:samplingrate_weight_results} presents the weight scores for various sampling rate candidates across different sensors. It reveals a trend where higher sampling rates are typically assigned greater weight scores.
However, the change speed in weight scores with decreasing sampling rates varies among sensors within the same task. Furthermore, the result indicates a correlation between a sensor’s sensitivity to sampling rate changes and its importance in the classification task, as illustrated in \cref{fig:sensor_weight_results} and \cref{fig:samplingrate_weight_results}.
To determine the most effective sampling rate combination for the sensors, we evaluated the classification performance sensitivity to changes in the sampling rates of each sensor. This classification result was obtained by maintaining the feature branch of the studied sampling rate while pruning branches of other sampling rates. 
The outcomes of this sampling rate optimization are depicted in \cref{fig:Sampling_optimization_result}.
The results demonstrate a significant performance decline when the sampling rate of the most critical sensor (identified by the highest weight score) is reduced. In contrast, sensors with lower importance scores exhibit less performance degradation. For instance, in the Opportunity-gesture task, only sensors $S4$ and $S9$ showed notable sensitivity to sampling rate changes, as detailed in Figure \cref{fig:Opportunity_gestures_samplingrate_performance}.
Thus, we can keep these two sensors at the highest sampling rate and the rest of the sensors at the lowest sampling rate to save both hardware and power consumption.
The classification performance in PAMAP2 and MHEALTH dataset has obvious degradation after the sampling rate of all sensors has been down to 60 Hz and 30 Hz accordingly as \cref{fig:PAMAP2_gestures_samplingrate_performance} and \cref{fig:MHEALTH_gestures_samplingrate_performance} shown.


Optimized sensors and their corresponding sampling rates were chosen based on a combined evaluation of weight score rankings, sensor importance, and the effects on classification performance using the progressive pruning method. \cref{tab:result_summary} summarizes these co-optimization results for both sensors and sampling rates. 
A notable finding is that the most efficient sensor optimization occurred with the MHEALTH dataset. In this case, five out of eight sensors were removed, leading to a 62\% reduction in model size, which only resulted in a minor performance decrease of approximately 3.19\% compared to the original model.
The results also show that for the PAMAP2 and MHEALTH datasets, performance impact was minimal even when the sampling rates for most retained sensors were lowered. Overall, the CoSS methodology for optimizing sensors and sampling rates successfully maintains classification performance while significantly reducing memory requirements (due to smaller model size), computational demands (owing to smaller kernel and window sizes), and power consumption (as a result of fewer and slower data transactions)


\section{Discussion}
With the CoSS framework, the sensor and sampling rate selection can be guided by ranking the weight scores of the sensor and sampling rate as well as the progressive pruned result, which demonstrated a more effective solution for selecting both sensor and sampling rate than an exhaustive search method.
In addition, we found that the weight scores of sensors and their sampling rates are closely related during the co-optimization process.
As shown in \cref{fig:relation_sensor_samplingrate}, the standard deviation was calculated from the weight scores of the sampling rate from each sensor to represent their sensitivity to the sampling rate, it can be found that the sensors with higher weight scores are more sensitive to the sampling rate in this work. 
Thus, assigning a higher sampling rate to these sensors with higher weight scores could be an optimal combination in multi-sensor-based HAR tasks. 
However, two major limitations of this work are also observed:
Firstly, the number of feature extraction branches could be very large if there are many sensors and sampling rate candidates, which leads to the training process being more time-consuming, although it just requires training once. 
Secondly, only the classification performance is considered in the framework during the sampling rate selection, which results in a higher sampling rate having a higher ranking; however, the hardware and computational cost of the CNN channel with different sampling rates is not considered, which should be integrated into the loss function and optimized during training in the future. 

\section{Conclusion}
In this work, we introduced the CoSS framework, designed to co-optimize sensors and sampling rates in HAR tasks for efficient utilization of sensor data. 
We proposed two types of trainable weight scores integrated into a feature-level fusion ANN. 
These scores assess the importance of both sensors and their sampling rates. 
By training the model once, we can rank these weight scores, guiding the progressive pruning of sensors and selecting the optimal sampling rate. 
To validate our framework, we tested its effectiveness in optimizing sensor modality and sampling rate selection on three public HAR benchmark datasets. 
The results show that the sensor and sampling rate combination selected via CoSS achieves similar classification performance to configurations using the highest sampling rate with all sensors but at a reduced hardware cost.
In future work, a neural network architecture search method will be considered to be integrated into the framework, by which both input complexity and neural network complexity could be reduced in the HAR task.

\section{Acknowledgments}
This work is supported by the European Union’s Horizon Europe research and innovation program (HORIZON-CL4-2021-HUMAN-01) through the "SustainML" project (grant agreement No 101070408).


\bibliography{aaai24}

\end{document}